\documentclass[twocolumn,prl,preprintnumbers,amsmath,amssymb]{revtex4}

\newcommand{\be}
{\begin{eqnarray}}

\newcommand{\ee}
{\end{eqnarray}}

\usepackage{graphicx} \usepackage{amsmath} \usepackage{dcolumn} \usepackage{bm} \usepackage{times}

\begin{document}

\bibliographystyle{unsrt}

\title{All-optical manipulation of electron spins in carbon nanotube quantum dots}

\author{Christophe Galland}  \author{Ata\c{c} Imamo\u{g}lu}
\affiliation{Institute of Quantum Electronics, ETH Zurich,
Wolfgang-Pauli-Strasse 16, CH-8093 Zurich, Switzerland}

\date{\today}

\begin{abstract}

We demonstrate theoretically that it is possible to manipulate electron or hole spins all optically
in semiconducting carbon nanotubes. The scheme we propose is based on the spin-orbit interaction
that was recently measured experimentally; we show that this interaction, together with an external
magnetic field, can be used to achieve optical electron-spin state preparation with fidelity
exceeding 99 $\%$. Our results also imply that it is possible to implement coherent spin rotation
and measurement using laser fields linearly polarized along the nanotube axis, as well as to
convert spin qubits into time-bin photonic qubits. We expect that our findings will open up new
avenues for exploring spin physics in one-dimensional systems.

\end{abstract}

\maketitle

Optical manipulation of spins in atoms or semiconductors relies on the presence of strong
spin-orbit interaction (SOI) either in the initial or final state of an optical transition. In
III-V semiconductors, it is the large spin-orbit splitting of the valence band states
\cite{Cardona} that enables efficient optical pumping of electron \cite{Atature} or nuclear spins
\cite{Lai,Gammon} and leads to a strong correlation between light helicity and electron spin
orientation \cite{Imamoglu}. In this context, one would argue that optical spin manipulation would
be hindered in semiconducting single-wall carbon nanotubes (CNTs): due to the weak spin-orbit
splitting in graphene \cite{Yao} and early experiments suggesting the presence of electron-hole
symmetry \cite{Jarillo}, it had been assumed that SOI in CNTs would be small for both electrons and
holes. In addition, the depolarization effect ensures that only electric fields linearly polarized
along the CNT axis couple strongly to electrons and holes \cite{Uryu,Lefebvre}, ruling out the
possibility of obtaining correlations between electron spin and photon polarization.

In this Letter, we describe a scheme for realizing efficient optical manipulation of spins in CNTs.
Our work is motivated by the recent experimental observation of SOI-induced zero-field spin
splitting in CNTs \cite{Kuemmeth}. The breakdown of electron-hole symmetry that is a consequence of
finite SOI implies that a finite external axial magnetic field could be used to cancel the
SOI-induced spin splitting of the hole, while retaining a finite splitting for the electron spin.
The presence of a magnetic field component perpendicular to the CNT axis then mixes the hole spin
states and allow for a very efficient spin-flip Raman coupling between the electron spin states. In
addition to analyzing the spin pumping efficiency as a function of the external magnetic and laser
fields, we discuss applications of the proposed scheme in quantum information processing.

The  band structure of a CNT can be derived from that of graphene, in which conduction and valence
bands are crossing at two inequivalent points in the reciprocal lattice (labeled K and K') with
linear dispersions. Since the K and K' points are at the boundaries of the first Brillouin zone ,
the states near the energy gap in semiconducting CNTs have a large azimuthal momentum $k_{\bot}$.
In a semi-classical picture, these states correspond to electrons having a fast clockwise or
counter-clockwise circular motion around the CNT's circumference, therefore exhibiting an orbital
magnetic moment $\mu_{orb}\approx 0.3\cdot d$[nm]~meV/T pointing along the axis, with opposite
signs for states originating from the two different valleys \cite{Kuemmeth,Minot}. The degeneracy
between these states is therefore lifted when a magnetic field $B_\parallel$ is applied along the
nanotube axis \cite{Minot,Srivastava}. The spin of electrons (or holes) also couples to the
magnetic field, yielding a Zeeman splitting $\Delta_Z = g \mu_B B_\parallel$, with $g\approx 2$ for
both electrons and holes \cite{Kuemmeth}. In this simple picture the electronic states should be
four-fold degenerate at zero magnetic field. Here, we focus on a CNT quantum dot (CNT QD) trapping
a single electron, since this is the system of primary interest from a quantum information
perspective \cite{excconf}.

\begin{figure}[t]
\includegraphics[width=8.3cm]{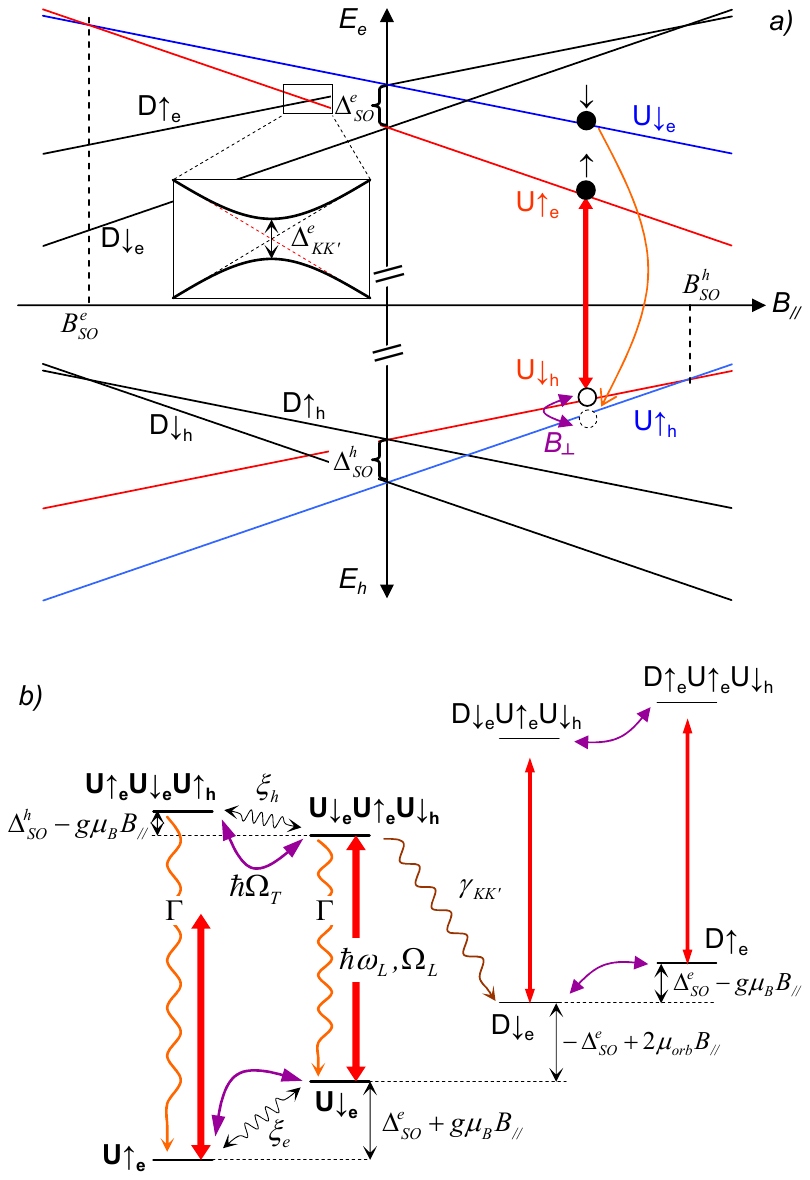}
\caption{ a) Energy diagram of the lowest electron (subscript \emph{e}) and hole (subscript
\emph{h}) states in a nanotube quantum dot as a function of the applied axial magnetic field
$B_{//}$ \cite{Kuemmeth,Bulaev}. b) Energy diagram of a singly charged nanotube quantum dot for
large $B_{//}$ showing the relevant optical transitions coupled by a laser field polarized along
the CNT axis with Rabi frequency $\Omega_L$. The orthogonal magnetic field causes a coherent
$\uparrow$-$\downarrow$ coupling with strength $\hbar \Omega_T = g \mu_B B_{\perp}$. The decay of
the excited states is assumed to be spin-conserving and mono-exponential with rate $\Gamma$. Spin
relaxation rates for electrons and holes are denoted by $\xi_e$ and $\xi_h$. K-K' mixing can cause
optically assisted valley-flip to the state D$\downarrow_e$ at an effective rate $\gamma_{KK'}$. }
\label{fig1}
\end{figure}

The SOI leads to a zero-field spin splitting $\Delta_{SO}$ between states with parallel and
anti-parallel spin and orbital magnetic moments \cite{Bulaev,Kuemmeth} (Fig.~1a). In the following
analysis, we rely on the experimentally measured values for $\Delta_{SO}$ \cite{Kuemmeth}. Since we
are interested in optical manipulation of spins, we consider a QD formed on a semiconducting CNT
with a diameter $d \sim 1.2$~nm having its lowest optical transition in the near-infrared ($\sim
1500$~nm) \cite{Weisman}. Extrapolating measured values from \cite{Minot, Kuemmeth} to this
diameter ($\mu_{orb}\propto d$ and $\Delta_{SO}\propto \frac{1}{d}$) we expect $\mu_{orb}\approx
0.36$~meV/T, $\Delta^e_{SO}\approx 1.5$~meV for electrons and $\Delta^h_{SO} \approx 0.9$~meV for
holes \cite{discrep}.

We first study the possibility of optical spin pumping using resonant laser fields. In Fig.~1a we
show the energy level diagram of the lowest electron and hole  states in a CNT QD under an axial
magnetic field $B_{\parallel}$ as confirmed experimentally by Kuemmeth \emph{et al.}
\cite{Kuemmeth}. We will label U (D) the states having a positive (negative) orbital magnetic
moment. These states originate from the two different valleys K and K' and mix very weakly in clean
CNTs ($\Delta_{KK'} \approx 65 \mu$eV in \cite{Kuemmeth}). Allowed optical transitions are of the
type U$\rightarrow$U or D$\rightarrow$D due to momentum conservation. The up and down arrows
represent the projection of the spin along the CNT axis ($\uparrow$ for $S_z = +\frac{\hbar}{2}$)
and the subscripts designate electron or hole states. We now apply an axial magnetic field
$B_{\parallel}$ and a laser field polarized linearly along the nanotube axis, strongly coupling
states from the same valley with opposite electron and hole spins. In addition, we assume the
presence of a magnetic field component $B_{\perp}$ orthogonal to the CNT axis that coherently mixes
the up and down electron (and hole) spin states.

Figure 1b shows an energy level diagram equivalent to that of Fig.~1a in the trion picture
\cite{Atature} where the four lowest energy spin states of a CNT QD as well as the optically
excited states with two electrons and one hole are depicted \cite{bright}. We choose the energy of
the laser to be resonant with the U$\downarrow_e
\rightarrow$U$\downarrow_e$U$\uparrow_e$U$\downarrow_h$ transition. The optically excited trion
state now couples to U$\downarrow_e$U$\uparrow_e$U$\uparrow_h$ because the hole spin precesses
around the perpendicular field $B_{\perp}$. Radiative recombination from
U$\downarrow_e$U$\uparrow_e$U$\uparrow_h$ leaves a spin-up electron U$\uparrow_e$. Since the
optical transition U$\uparrow_e \rightarrow$U$\downarrow_e$U$\uparrow_e$U$\uparrow_h$ is detuned by
$\Delta^e_{SO}+\Delta^h_{SO}$ from the applied laser field, light scattering (experimentally
measured by laser absorption) will vanish \cite{Atature}, ensuring that the resident electron spin
remains in state U$\uparrow_e$. Conversely, preparation of the spin in the U$\downarrow_e$ state
can be achieved by tuning the laser field onto resonance with the U$\uparrow_e \rightarrow$
U$\downarrow_e$U$\uparrow_e$U$\uparrow_h$ transition.

To assess the efficiency of optical spin pumping as a function of the applied magnetic  and laser
fields we have performed numerical simulations using the optical Bloch equations for the 4-level
system shown in the left part of Fig.~1b. We ignore K-K' mixing for the time being. The spin-flip
rates $\xi_e$ ($\xi_h$) for electrons (holes) are dominated by phonon-assisted spin relaxation
\cite{Bulaev,Borysenko}: these rates are expected to have magnitudes varying from $1$~$\mu$s$^{-1}$
to 1 ms$^{-1}$. We take the exciton recombination time to be $\Gamma^{-1} \approx 40$~ps,
corresponding to the photoluminescence (PL) lifetime measured on individual CNTs
\cite{Hoegele,Galland}. The narrowest reported nanotube PL linewidths ($0.25 - 0.5$~meV
\cite{Srivastava,Htoon}) however, are an order of magnitude larger than the lifetime broadening. We
therefore include a Markovian dephasing rate of both optical transitions
$\hbar\gamma_{deph}=0.25$~meV.

The resulting spin population  imbalance as a function of the axial and perpendicular components of
the magnetic field is shown in Fig.~2a, when the laser is resonant with the red (lowest energy)
transition. First, we remark that spin preparation with a fidelity close to 1 is possible at almost
any axial field provided that the perpendicular field is of the order of a few 100 mT. But the main
feature revealed by the simulation is the peculiar behavior of the system when the Zeeman splitting
caused by $B_{\parallel}$ cancels the spin-orbit splitting for either electrons or holes: when
$B_{\parallel}=B^h_{SO} = \Delta^h_{SO}/(g \mu_B)$, the states
U$\downarrow_e$U$\uparrow_e$U$\downarrow_h$ and U$\downarrow_e$U$\uparrow_e$U$\uparrow_h$ have the
same energy; as a consequence the mixing induced by even a vanishingly small $B_{\perp}$ suffices
to yield very efficient electron spin pumping. On the contrary, for $B_{\parallel}=B^e_{SO} =
\Delta^e_{SO}/(g \mu_B)$, the resonance occurs between the electronic states and the electron spin
remains randomized for all values of $B_{\perp}$.

\begin{figure}[t]
\includegraphics[width=8.1cm]{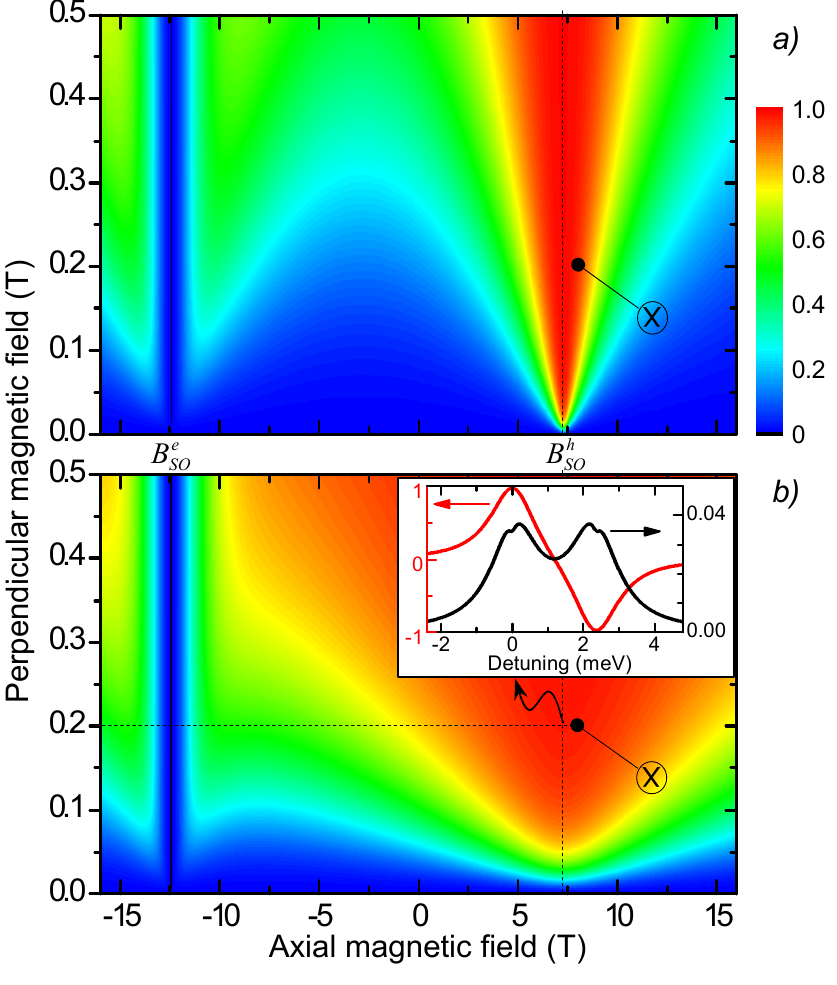}
\caption{ The contour plot of  the population difference between states U$\uparrow_e$ and
U$\downarrow_e$ as a function of the magnetic field components when the laser is kept resonant with
the U$\downarrow_e \rightarrow$U$\downarrow_e$U$\uparrow_e$U$\downarrow_h$ transition (zero of the
detuning scale in the inset). We take $\xi_e = \xi_h = 0.1$ $ \mu$s$^{-1}$ and $\Gamma^{-1} =
40$~ps. In (a), we assume that the optical transitions are broadened by Markovian dephasing with
rate $\hbar\gamma_{deph}=0.25$~meV. In (b), we consider a scenario where the coherence between
states U$\downarrow_e$U$\uparrow_e$U$\downarrow_h$ and U$\downarrow_e$U$\uparrow_e$U$\uparrow_h$
also undergoes fast dephasing at the same rate ($0.25$~meV). In the inset to part (b), we show the
laser absorption (black line) and  population difference (red line) between U$\uparrow_e$ and
U$\downarrow_e$ states for external magnetic fields $B_{\parallel}=B^h_{SO}$ and $B_{\perp}=0.2$~T.
The absorption (right scale) is normalized to its maximum value in absence of spin pumping (when
$B_{\perp}=0$~T).
}\label{fig2}
\end{figure}

Since it is likely that the excited trion states undergo  faster dephasing than the single electron
ground states, we performed an additional simulation where the coherence between the two trion
states is also dephased at the rate $\hbar\gamma_{deph}$. The result, shown in Fig.~2b, is
qualitatively similar to the case depicted in Fig.~2a. For $B_{\parallel} \approx B^h_{SO}$ and
$B_{\perp} = 0.2$~T (inset Fig.~2b) very efficient spin preparation in either state U$\uparrow_e$
or U$\downarrow_e$ is achieved upon tuning the laser across the U$\downarrow_e
\rightarrow$U$\downarrow_e$U$\uparrow_e$U$\downarrow_h$ and U$\uparrow_e
\rightarrow$U$\downarrow_e$U$\uparrow_e$U$\uparrow_h$ transitions. Pauli blockade leads to a drop
in absorption by more than an order of magnitude: this would be the experimental signature of spin
pumping in differential transmission measurements \cite{Atature}. Our results demonstrate that
selective optical spin preparation in CNT QDs is in experimental reach. A limitation would however
appear for very small QDs ($\sim 10$~nm): it was shown recently that the ohmic coupling of strongly
confined excitons to one-dimensional acoustic phonons in CNTs leads to asymmetric absorption
spectrum with a pronounced blue tail, extending over a few meV \cite{Galland}. This pure-dephasing
process would nevertheless not alter the efficiency of spin pumping when driving the lower-energy
resonance with a red-detuned laser.

The presence of K-K' valley mixing  (characterized by the splitting $g\mu_B B_{KK'}=\Delta_{KK'}$)
will result in a finite probability that the electron spin leaves the Hilbert space spanned by
U$\downarrow_e$ and U$\uparrow_e$.  We denote the effective spin-flip Raman scattering rate from
state U$\downarrow_e$ to U$\uparrow_e$ with $\gamma_{\uparrow\downarrow}$ and the effective rate
for a laser assisted transition from U$\downarrow_e$ to D$\downarrow_e$ with $\gamma_{KK'}$. Using
rate equations we obtain: $ \frac{\gamma_{\uparrow\downarrow}}{\gamma_{KK'}} \approx
\frac{B_{\perp}^2/(\Delta^h_{SO}-g\mu_B B_{\parallel})^2}{B_{KK'}^2/(
\Delta^e_{SO}-2\mu_{orb}B_{\parallel})^2} $ which is maximum when $B_{\parallel} \rightarrow
B^h_{SO}$. We find that  using the experimentally measured parameters, efficient spin pumping is
possible in large regions of magnetic fields. For example at point X in Fig.~2, for which
$B_{\parallel} = 8$~T $\approx B^h_{SO} + 0.75$~T and $B_{\perp} = 0.2$~T, we have
$\frac{\gamma_{\uparrow\downarrow}}{\gamma_{KK'}} > 100$. Once the system goes through a
valley-flip Raman scattering to state D$\downarrow_e$, the applied laser field will be detuned from
the transition to the state D$\downarrow_e$U$\uparrow_e$U$\downarrow_h$, due to the exchange terms
of Coulomb interaction (see Fig.~1b, right). If $B_{KK'}$ is large it would therefore be necessary
to use a second re-pumping laser on this transition to reintroduce the electron to the U-valley.

In most experiments it has been observed that the lifetime of excitons is more than an order of
magnitude shorter than the predicted radiative lifetime \cite{Hoegele,Galland,Hirori}. While
radiative broadening can be enforced by embedding CNTs in cavity structures with a large Purcell
factor \cite{Badolato}, understanding the nature of non-radiative relaxation is crucial for
identifying the limits of optical spin manipulation. In particular, if this relaxation is not
spin-conserving, then spin pumping becomes efficient for an even larger range of applied magnetic
field strengths. Most probable mechanisms for fast non-radiative decay proposed so far are
phonon-assisted relaxation and/or multi-particle Auger processes \cite{Perebeinos}. Since these
processes are spin-conserving, they will not alter the efficiency of spin pumping.

Having demonstrated that it is viable to  prepare a single spin optically, we turn to coherent spin
rotation and spin measurement. By using two laser fields satisfying two-photon Raman resonance
condition under the same external magnetic field configurations that allow for efficient spin
pumping, we can implement deterministic spin rotation \cite{Imamoglu}. To realize all-optical spin
measurement, the field $B_{\perp}$ mixing the electron (hole) spin states must be turned off. In
this limit, presence or absence of light scattering (or absorption) upon excitation by a resonant
laser conveys information about the spin state \cite{Imamoglu}. For spin measurements, minimizing
spin-flip non-radiative relaxation and inter-valley scattering is crucial. We also point out that
all of our results would apply for a single-hole charged CNT QD as well.

Next, we address the possibility of transferring quantum information stored in the CNT QD electron
spin to a generated photon. Given that the polarization of the photon is fixed by the geometry, the
logical choice is to use time-bin entanglement \cite{Brendel}. We assume that our CNT QD is coupled
to an optical cavity whose energy $\omega_{cav}$ is resonant with the transition U$\downarrow_e
\rightarrow$U$\downarrow_e$U$\uparrow_e$U$\downarrow_h$ (Fig.~3). Using combinations of laser
pulses one can prepare an initial state in the coherent spin superposition: $|\psi_{in} \rangle =
(\alpha | $U$\uparrow_e \rangle + \beta | $U$\downarrow_e \rangle)\otimes | 0_c \rangle$ where $|
0_c \rangle$ is the empty cavity mode. We now send two well separated $\pi$-pulses at time $t_1$
and $t_2$ with respective energies $\omega_a$ and $\omega_b$ as shown in Fig.~3. We define the two
creation operators $a^\dagger$ ($b^\dagger$) for cavity-mode photons emitted immediately after
pulse 1 (pulse 2). The optical transition at frequency $\omega_b$ is allowed because of the mixing
induced by $B_{\perp}$, and the rates of both transitions can be made identical by adjusting the
pulse intensities. The first pulse excites the trion state if and only if the spin is initially
down. In this case Purcell effect ensures very fast spontaneous emission and projection onto the
state $| $U$\downarrow_e \rangle \otimes a^\dagger | 0_c \rangle$. If the spin is initially up, the
transition is Pauli-blocked and we are left with $| $U$\uparrow_e \rangle \otimes | 0_c \rangle$.
The initial state has thus evolved to: $ |\psi_{1} \rangle = \alpha | $U$\uparrow_e \rangle \otimes
| 0_c \rangle + \beta | $U$\downarrow_e \rangle \otimes a^\dagger | 0_c \rangle $. We can do the
same analysis for the second pulse and find that the final state is: $ |\psi_{f} \rangle = |
$U$\downarrow_e \rangle \otimes (\alpha b^\dagger + \beta a^\dagger) \otimes | 0_c \rangle $ where
quantum information has been mapped onto a photon time-bin qubit. We emphasize that time-bin qubits
are promising candidates for long range quantum communication using optical fibers \cite{Marcikic}
and that CNTs can be chosen to emit in the desired wavelength window.

\begin{figure}[h]
\includegraphics[width=8.4cm]{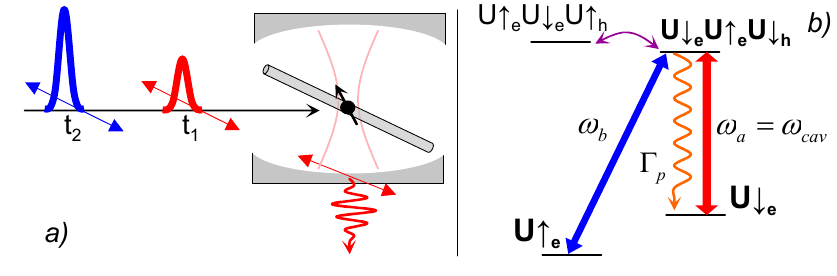}
\caption{ a) Schematics of the cavity QED setup discussed in the text. Both laser pulses are
polarized along the nanotube axis. Provided that the pulse separation $t_2 - t_1$ is larger than
all the other time scales (inverse Rabi frequency and cavity-enhanced exciton decay rate) quantum
information can be efficiently encoded in a photon time-bin qubit. b) Energy diagram of the
nanotube quantum dot with the relevant transitions and rates used in the scheme.} \label{fig3}
\end{figure}

We remark that one of the most interesting perspectives enabled by the considerations of this
letter is the study of nuclear spin physics. The possibility of electron spin pumping should allow
for the optical manipulation of nuclear spin ensembles, which has been successfully achieved in
GaAs-based structures \cite{Lai,Gammon}. However, experimental knowledge of the strength and
characteristics of hyperfine interaction in CNTs is still lacking. Of particular interest in this
context would be dynamic nuclear spin polarization in a CNT QD where hundreds or thousands of
$^{13}$C atoms would form an ideal $I=\frac{1}{2}$ spin bath. Alternatively, using high-purity
$^{12}$C CNTs, one may realize QDs interacting with only 1 or 2 nuclear spins \cite{Childress}.

This work was supported by a grant from the Swiss National Science Foundation (SNSF).


\begin{thebibliography}{}
\bibitem{excconf} Our scheme requires that the charged
exciton (trion) be trapped as well. We note that the recent experiments demonstrated strong
unintentional confinement of nanotube excitons \cite{Hoegele,Galland}, possibly around charged
defects. Enhanced interactions in 1D systems should lead to a trion binding energy well exceeding
several meVs, suggesting that external gates could confine trions as well as electrons, even when a
neutral exciton is not confined. Alternatively, local strain was shown to shift the excitonic
states very efficiently \cite{Maki,Yang} and could be used for trion localization.
\bibitem{discrep} The discrepancy between $\Delta^e_{SO}$ and $\Delta^h_{SO}$ found in \cite{Kuemmeth} is theoretically unexplained and seemingly dependent on experimental conditions.It is neither
significant nor necessary for the scheme that we propose.
\end{thebibliography}
\end{document}